\documentclass[letter]{aa} 

\usepackage{graphicx}
\usepackage{txfonts}
\usepackage[T1]{fontenc}
\usepackage{ae,aecompl}
\usepackage[colorlinks=true,linkcolor=blue,citecolor=blue,urlcolor=blue]{hyperref}

\usepackage{amssymb} 
\usepackage{footnote}
\usepackage{adjustbox}
\usepackage{lipsum}
\usepackage{mwe}
\usepackage{verbatim}
\usepackage{blindtext}
\usepackage{mathtools}
\usepackage{relsize}
\usepackage[group-separator={,},group-minimum-digits=4]{siunitx}
\usepackage{rotating}
\usepackage{adjustbox}
\usepackage{multirow}
\usepackage{soul}
\usepackage{placeins}

\newcommand{\Msun}{${\rm M_\odot}$}
\newcommand{\Lsun}{${\rm L_\odot}$}
\newcommand{\Rsun}{${\rm R_\odot}$}

\newcommand{\Teff}{${\rm T_{eff}}$} 

\newcommand{\Lbol}{${\rm L_{bol}}$}

\newcommand{\kms}{\,km\,s$^{-1}$}

\begin{document} 

   \title{Discovery of a hot post-AGB star in Galactic globular cluster E3}
   \titlerunning{Discovery of hot PAGB star in E3}


   \author{R. Kumar
          \inst{\ref{inst1},\ref{inst2}}, 
          A. Moharana\inst{\ref{inst3}}, 
          S. Piridi\inst{\ref{inst2}},
          A. C. Pradhan\inst{\ref{inst2}}, 
          K.G. Hełminiak\inst{\ref{inst3}},
          N. Ikonnikova\inst{\ref{inst4}},
          A. Dodin\inst{\ref{inst4}},
          R. Szczerba\inst{\ref{inst5},\ref{inst3}},
          M. Giersz\inst{\ref{inst6}},
          D. K. Ojha\inst{\ref{inst7}},
          \and
          M. R. Samal\inst{\ref{inst1}}
          }
    \authorrunning{R. Kumar et al.}

   \institute{Astronomy and Astrophysics Division, Physical Research Laboratory, Navrangpura, Ahmedabad - 380009, Gujarat, India \label{inst1} \\ \email{ranjankmr488@gmail.com, acp.phy@gmail.com}
         \and
             Department of Physics and Astronomy, National Institute of Technology, Rourkela - 769008, Odisha, India \label{inst2}
        \and
             Nicolaus Copernicus Astronomical Center, Polish Academy of Sciences, ul. Rabiańska 8, 87-100 Toruń, Poland\label{inst3}
        \and
             Sternberg Astronomical Institute of the Lomonosov Moscow State University, University av. 13, 119234 Moscow, Russia\label{inst4}
        \and
            Xinjiang Astronomical Observatory, Chinese Academy of Sciences 150 Science 1-Street, Urumqi, Xinjiang 830011, China\label{inst5}
        \and
             Nicolaus Copernicus Astronomical Centre, Polish Academy of Sciences, ul. Bartycka 18, 00-716 Warsaw, Poland\label{inst6}
        \and
             Department of Astronomy and Astrophysics, Tata Institute of Fundamental Research, Mumbai - 400005, India\label{inst7}
             }
             
   \date{Received: 28 February 2024 / Accepted: 20 April 2024}


\abstract
{
We report a new hot post-asymptotic giant branch (PAGB) star in the Galactic globular cluster (GC) E3, which is one of the first of the identified PAGB stars in a GC to show a binary signature. The star stands out as the brightest source in E3 in the \mbox{{\em Astrosat}}/UVIT images. We confirmed its membership with the cluster E3 using Gaia DR3 kinematics and parallax measurements. We supplemented the photometric observations with radial velocities (RVs) from high-resolution spectroscopic observations at two epochs and with ground- and space-based photometric observations from 0.13 $\mu$m to 22 $\mu$m.  We find that the RVs vary over $\sim$6 \kms\ between the two epochs. This is an indication of the star being in a binary orbit. A simulation of possible binary systems with the observed RVs suggests a binary period of either 39.12 days or 17.83 days with mass ratio q$\geq$1.0. The [Fe/H] derived using the high-resolution spectra is $\sim -$0.7 dex, which closely matches the cluster metallicity. The spectroscopic and photometric measurements suggest \Teff\ and  $\log g$ of the star as 17\,500$\pm$1\,000~K and 2.37$\pm$0.20~dex, respectively. Various PAGB evolutionary tracks on the Hertzsprung--Russell (H-R) diagram suggest a current mass of the star in the range 0.51$-$0.55 \Msun. The star is enriched with C and O abundances, showing similar CNO abundances compared to the other PAGB stars in GCs with the evidence of the third dredge-up on the AGB phase. 
}

   \keywords{Stars: AGB and post-AGB --
                (Galaxy:) globular clusters: individual: E3 
               }

   \maketitle
%

\section{Introduction}  \label{sec:introduction}

Post-asymptotic giant branch (PAGB) stars are the most luminous objects in globular clusters (GCs). They evolve from the top of the AGB and move leftward in the HR diagram at almost constant luminosity \citep{Kwok1982, Kwok1993}. They are rare in GCs (one or two per cluster is expected) due to the fact that PAGB stars have a very short lifetime \citep[${\rm \leq 0.5\ Myr; }$][]{Moehler2019} compared to the age of GCs \citep[10$-$12~Gyr;][]{Harris1996}. Nevertheless, a handful of PAGB stars have been detected in more than 150 Galactic GCs \citep[for details, see][]{Moehler2019, Davis2021, Dixon2024}. 

There are three major categories of PAGB stars observed in GCs based upon the evolution of their progenitors on the horizontal branch (HB), the core He burning phase. The first category is for HB stars with H-rich envelopes, which ascend towards the AGB; they have several thermal pulses at the tip of the AGB. Their post-AGB phase is treated as the normal PAGB phase \citep{Greggio1990, Dorman1993, Moehler2019}. The second category is for  HB stars with H-poor envelopes but H-envelopes > 0.02 \Msun, which ascend towards the AGB, but never reach the AGB tip and leave the AGB phase early\footnote{Either having one or two thermal pulses or without experiencing any thermal pulse.} \citep{Moehler2019}.  Their PAGB phase is known as post-early-AGB (PEAGB) phase \citep{Greggio1990, Dorman1993, Moehler2019}. They are less luminous than the normal PAGB stars \citep{Dorman1993, Moehler2019}. The third category of PAGB stars is for stars that evolved from HB stars with H-envelopes < 0.02 \Msun\ (extreme-HB stars) \citep{Dorman1993, Moehler2019}. They never reach the AGB phase; rather, they remain hot and luminous during their post-HB evolution, and ultimately cool down through the white dwarf (WD) phase. These PAGB stars are known as AGB-manqu\'e stars \citep{Dorman1993, Gratton2010, Moehler2019}.  

In a cluster with a large binary fraction, a few red-giant-branch (RGB) stars in a binary system lose enough mass to their companion. They become inefficient in igniting He during the RGB phase and evolve off the RGB at the luminosity comparable to their RGB tip. These stars are called post-RGB stars \citep{Lei2015, Kamath2016}. They have similar luminosities to PEAGB stars and ignite He either on their post-RGB phase or the WD cooling track \citep{Cassisi2003, Bertolami2008, Lei2015}.

\citet{Moehler2019} performed a comprehensive analysis of 78 Galactic GCs to identify the hot PAGB stars. They found 19 new PAGB stars in addition to 12 previously known PAGB/PEAGB stars. \citet{Prabhu2021} have detected five PAGB stars in the \object{NGC\,2808} using ultraviolet (UV) images of Ultraviolet Imaging Telescope (UVIT) on board {\em AstroSat} \citep{Tandon2017}. \citet{Davis2021} studied 97 Galactic GCs using optical filters and found 13 cooler (F and G type) PAGB stars among them. \citet{Dixon2024} studied the  CNO and s-process abundances of 17 previously known  PAGB stars in the Galactic GCs to find evidence of the third dredge-up (3DU) among GC PAGB stars. Considering all the above surveys, a total of 42 unique PAGB/PEAGB/AGB-manqu\'e stars have been observed so far only in $\sim$20 GCs when exploring more than 150 Galactic GCs. None of them has shown a binary signature (neither photometric nor spectroscopic). However, the low-mass PAGB stars, V29 in $\omega$ cen and K648 in M15, indicate that they have evolved in a binary system \citep{Dixon2024}.

In this paper we report the discovery of a new hot PAGB star in Galactic GC E3 \citep[\object{ESO\,37-1};][]{Lauberts1976} using UV images from UVIT and Gaia DR3 data. Cluster E3 \citep[RAJ2000=09:20:57.07, DEJ2000=$-$77:16:54.8;][]{Harris1996, Harris2010} is one of the faintest GCs in the Milky Way, and is likely to be the least massive one \citep[${\rm 2.9 \times 10^3\ M_{\odot}}$;][]{Sollima2017}. It is a metal-rich cluster \citep[${\rm [Fe/H] \sim -0.7\ dex}$;][]{Marcos2015, Salinas2015, Monaco2018} with a moderate reddening \citep[${\rm E(B-V)=0.29}$;][]{Schlafly2011}. The cluster has the largest binary fraction (72\%) among Galactic GCs \citep{Milone2012}. 

\section{Photometric and spectroscopic observations} \label{sec: observation}
 
\begin{figure}
    \centering
    \includegraphics[width=0.95\columnwidth]{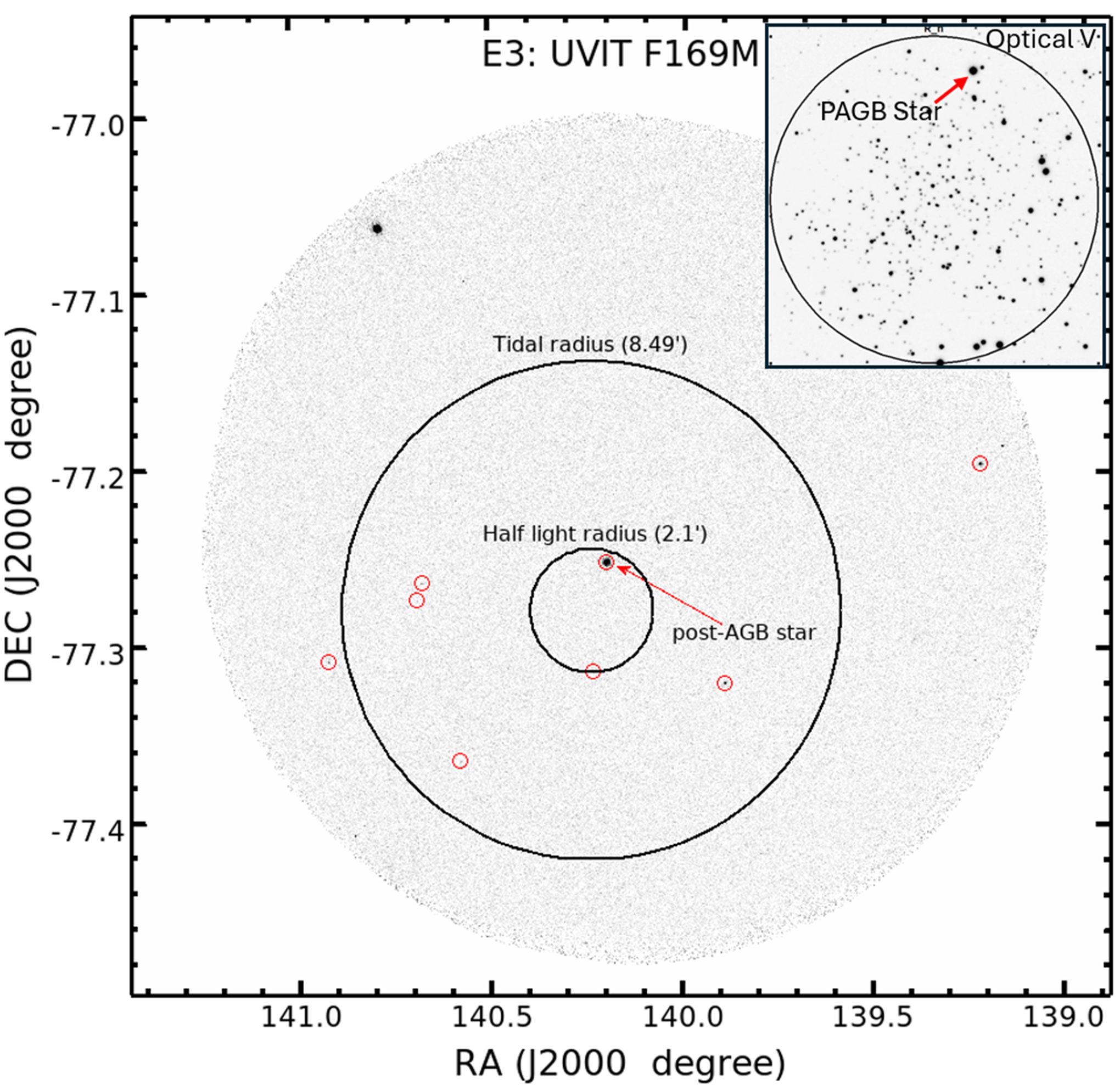}
    \caption{UVIT F169M filter image of E3. The half-light and tidal radius of E3 \citep[${\rm R_h=2.1'\ and\  R_t=8.49',}$][]{Harris1996, Harris2010} are well within the field of view of UVIT. The red arrow indicates the probable post-AGB star. In the inset (upper right corner)   the optical V-band image \citep{Stetson2019} of the cluster (up to half-light radius) is shown. The PAGB star is easily distinguishable as one of the cluster's brightest stars in both images.}
    \label{fig:baf2_mage}
\end{figure}

Cluster E3 was observed in two far-UV (FUV: 1300$-$1800 \AA) filters, F154W and F169M, of UVIT. The total exposure times of the  F169M and F154W filters are 6047 sec and 3546 sec, respectively. The field of view (FoV) of UVIT is sufficient   to cover the entire cluster \citep[tidal radius = 8.49 arcmin,][]{Harris1996}. However, we see only two FUV bright sources within the  FoV of UVIT; one is within the half-light radius of the cluster, and the other is at the edge of the image (Fig.~\ref{fig:baf2_mage}). The sources at the edges of the UVIT FoV are unreliable due to edge artefacts; hence, they were excluded during photometry. Finally, we detected eight sources in the FUV (red circles in Fig.~\ref{fig:baf2_mage}). 

\begin{figure*}
 \centering
    \includegraphics[width=0.325\textwidth]{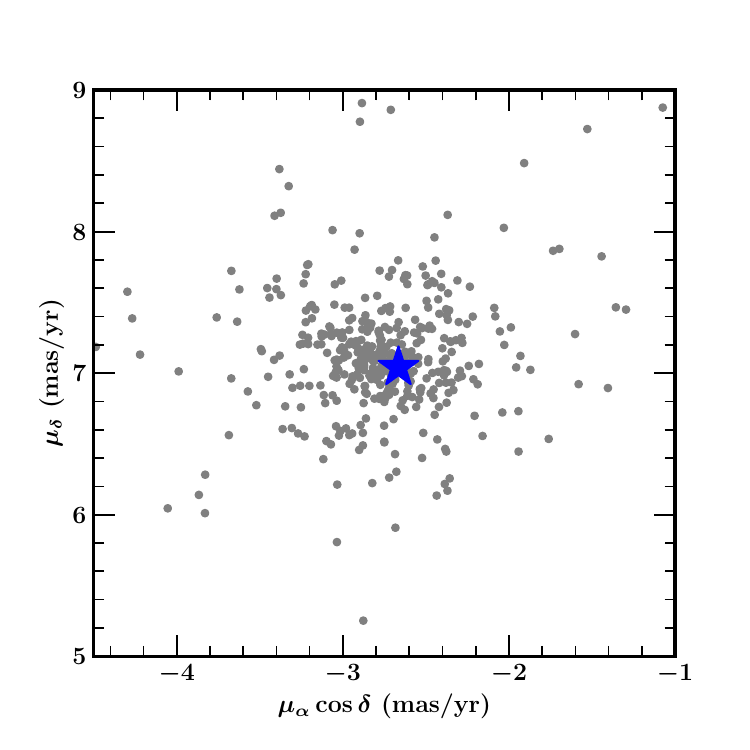}
    \includegraphics[width=0.325\textwidth]{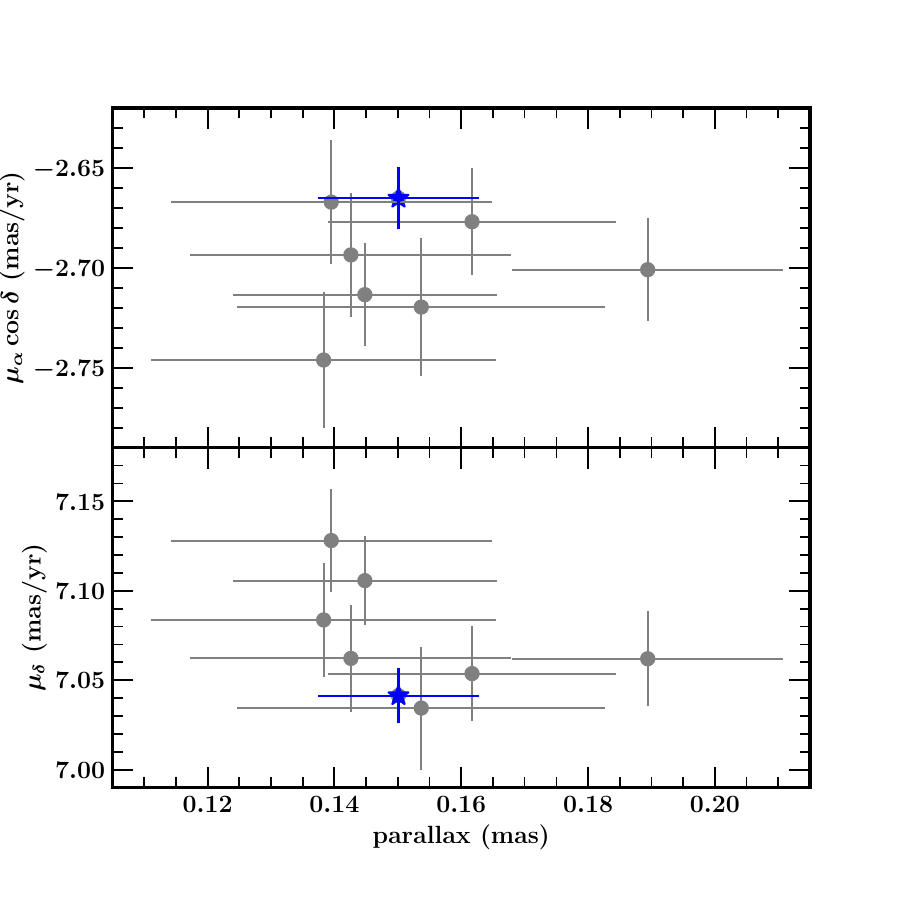}
    \includegraphics[width=0.325\textwidth]{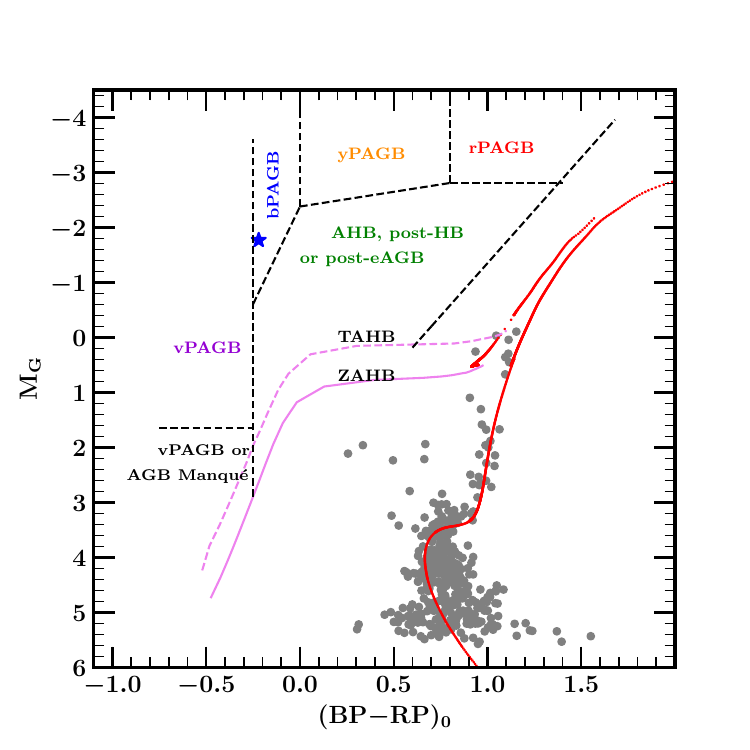}
    \caption{ Diagrams showing the kinematics and photometric analysis of the cluster member sources with the Gaia data. The left panel shows the vector-point diagram of the cluster member sources from the Gaia catalogue \citep[grey dots]{Vasiliev2021} and the UVIT observed cluster member source (blue asterisk). The middle panel shows the parallax vs proper motion in RA (top) and DEC (bottom) for the Gaia-observed cluster members having parallax S/N (plx/e\_plx) $>$ 5 (grey dots) and the UVIT observed cluster member source (blue asterisks). The right panel shows BP$-$RP versus M$_{\rm G}$ CMD in the absolute magnitude plane. The Gaia G magnitudes (grey dots) are scaled to absolute magnitudes using distance modulus, m$-$M = 14.50 \citep{Baumgardt2021}. The BaSTI-IAC isochrone of age 11 Gyr, [Fe/H] = $-0.70$ dex is overplotted as a red solid line. The zero-age HB (ZAHB) and terminal-age HB (TAHB) loci are shown as solid pink and dashed lines, respectively. The black dashed line shows different subcategories of PAGB and post-HB stars, as defined in \citet{Bond2021}. }
    \label{fig:gaia_pm}
\end{figure*}

\begin{table*}
    \centering
    \caption{Observation details of the PAGB star.}
    \adjustbox{max width=\textwidth}{
    \begin{tabular}{ccccccccc}
    \hline \hline
      RA (J2000) &  DEC (J2000) & Parallax & PMRA & PMDEC & Gaia/G  & UVIT/F154W & UVIT/F169M & Gaia ID \\
      (hh:mm:ss.ss) & (dd:mm:ss.ss) & (mas) & (mas/yr) & (mas/yr) & & & &  \\
      \hline
      09:20:48.02  & -77:15:15.36 & 0.1501 $\pm$ 0.0127 & -2.6651 $\pm$ 0.0155 & 7.0414 $\pm$ 0.0155 & 13.576 & 14.700 & 14.670 & 5203319520096719360 \\
      \hline
    \end{tabular}
    }
    \label{tab:pAGB_details}
\end{table*}

We used the GC catalogue of Gaia DR3 (GCG21; \citealt{Vasiliev2021}) for cluster membership analysis (see details of the analysis in Appendix~\ref{appendix: gaia}). We found only one out of eight UVIT observed sources to be a cluster member, which is indicated by the red arrow in Fig.~\ref{fig:baf2_mage}. The source is bright in FUV and optical bands (Table \ref{tab:pAGB_details}) and lies well within the half-light radius of the cluster. In Fig.~\ref{fig:gaia_pm} we show all the Gaia observed cluster member sources (grey dots) and the UVIT observed cluster member source (blue asterisk) on the vector-point diagram (left panel) and the parallax versus PMRA (middle upper panel) and parallax versus PMDEC (middle lower panel). We find that the proper motion and the parallax values of the UVIT observed source are in good agreement with the Gaia observed cluster member sources (Fig.~\ref{fig:gaia_pm}). The Gaia BP$-$RP versus  G colour-magnitude diagram (CMD) (right panel of Fig.~\ref{fig:gaia_pm}) suggests that the source is an evolved PAGB star in the GC E3. 

\subsection{Spectroscopic observations of the PAGB star}

The spectroscopic observations of the PAGB star were obtained at two epochs (17 February 2023 and 26 March 2023) with a high-resolution (R$\sim$28\,000) {\mbox CHIRON}\footnote{\href{http://www.astro.yale.edu/smarts/1.5m.html}{http://www.astro.yale.edu/smarts/1.5m.html}} spectrograph \citep{Tokovinin2013} installed on the SMARTS 1.5 m telescope at the  CTIO observatory, Chile. The observation details are given in Table \ref{tab:pagb_spec}, and the reduction procedure of the spectra is explained in Appendix~\ref{appendix: A}.

\subsubsection{Radial velocities}

We measured the radial velocities (RVs) using our own implementation \citep{Helminiak2019} of the {\sc todcor} technique \citep{Zuckermazeh1994}. The procedure is described in Appendix~\ref{appendix: B}. The barycentric RVs were estimated as 40.20$\pm$0.57 \kms\ and 34.66$\pm$0.85 \kms\ for the spectra observed at the first and second epochs, respectively. These values are quite high with respect to the mean RV of the cluster \citep[12.6 \kms;][]{Monaco2018}. However, the change in velocities between two epochs is quite substantial; this signals the possible binarity of the star. Since the cluster has a large binary fraction \citep{Milone2012}, it is possible that the identified PAGB star is evolving in a binary system. In Appendix \ref{appendix: C} we present the performed simulations for possible binary configurations of the PAGB star with constraints from the RVs. The simulations create binary models with different primary mass values in the range 0.2$-$0.8 \Msun. We find the period of the binary models is spread around 39.12 days (51\% of the models) and 17.83 days (16.4\% of the models) (Fig.~\ref{fig:period_q}). This suggests a close orbit of the system. The mass ratio\footnote{The ratio of the companion mass to the PAGB mass.} (q) of the system lies in the range of 0.2$-$1.5. \cite{Milone2012} calculated photometric binary fractions for E3 and found that around 40\% of the cluster members have a binary with q>0.5. We do not see another companion in the spectra, so q>1.5 (a conservative upper limit) is unlikely. Since the source is close to the half-light radius, it is likely to have a heavier companion. Hence, we suggest q$\geq$1.0 for the binary system. However, further spectroscopic monitoring is required to confirm the binarity and characterise the companion.

\subsubsection{Atmospheric parameters and abundances} 

The atmospheric parameters (\Teff\ and $\log g$) of the PAGB star were derived from CHIRON spectra using a relation involving the ionisation balance between \ion{Si}{ii} and \ion{Si}{iii}, for a set of \Teff\ and $\log g$ pairs from TLUSTY model grid BSTAR2006 \citep{Lanz2007}. The procedure of atmospheric parameter estimation is described in Appendix~\ref{appendix: D}. We find T$_{\rm eff}=17\,000_{-700}^{+400}$~K, $\log g = 2.37\pm0.20$, and $\log \varepsilon_{\rm Si}=7.31\pm0.05$ from the spectral analysis of the PAGB star.

Using the above parameters derived from the spectra of the PAGB star, we derived abundances ($\varepsilon$) of all elements detected in the spectrum. The chemical composition analysis of the individual elements is given in Appendix~\ref{appendix: E}, and the abundance value of each element is given in Table \ref{tab:abund}. The derived individual abundances of metals in the star are two to five times lower than in the Sun (Table\,\ref{tab:abund}), suggesting that the star is metal poor. The [Fe/H] abundance ($-$0.7$\pm$0.2 dex) matches  the cluster metallicity \citep{Monaco2018}. Unlike this star, many PAGB binaries   show depletion of iron \citep{vanWinckel2003}. However, in a recent study to understand depletion in binary RV Tauri stars (and a few binary PAGB stars), \cite{Gezer2015} found that iron depletion is prevalent in binaries with an accretion disk or with evidence of a former disk. The binaries without a disk do not show any depletion. Since we do not find any signature of an accretion disk (or depletion),  the identified PAGB star might be a case where the evolution in a binary does not involve the formation of an accretion disk.

\subsection{Spectral energy distribution of the PAGB star}

We performed the spectral energy distribution (SED) fitting of the PAGB star in the VO SED analyzer \citep[VOSA; ][]{Bayo2008}. The SED fitting procedure is described in Appendix~\ref{appendix: F}. We found that the best-fitting model grid has \Teff=18\,000~K and $\log g$ = 2.25. The best-fitting model and observed fluxes are shown in Fig.~\ref{fig:sed_fit}. VOSA also provides the bolometric luminosity (\Lbol) and radius of the star (R) from their best-fit stellar atmosphere model grid based on the distance of the star and the total observed fluxes from the UV to IR bands. We found \Lbol\ and $R$ of the identified PAGB star as 2136 $\pm$ 150 \Lsun\ (${\rm \log (L/L_{\odot}) = 3.33\pm 0.03}$) and 4.61$\pm$0.14 \Rsun, respectively. 

\section{Evolutionary status of the PAGB star} \label{sec: evolutionary status}

\begin{figure*}
    \centering
    \includegraphics[width=0.64\textwidth]{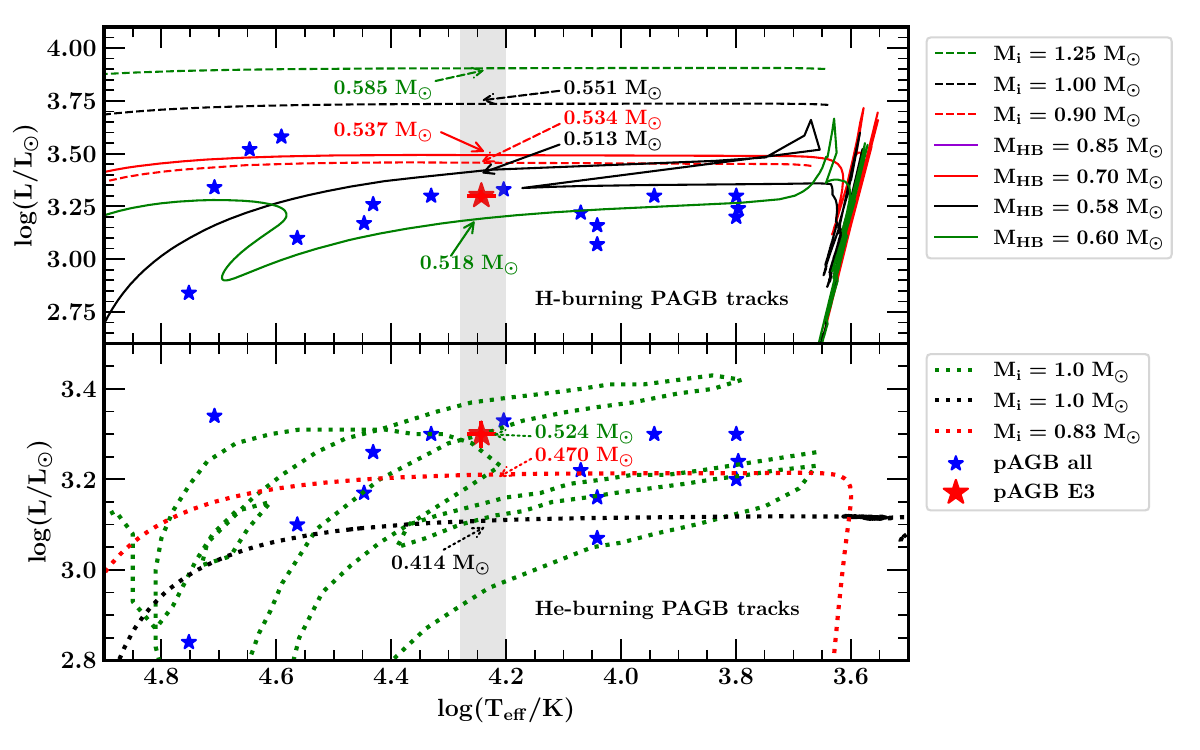}
    \includegraphics[width=0.35\textwidth]{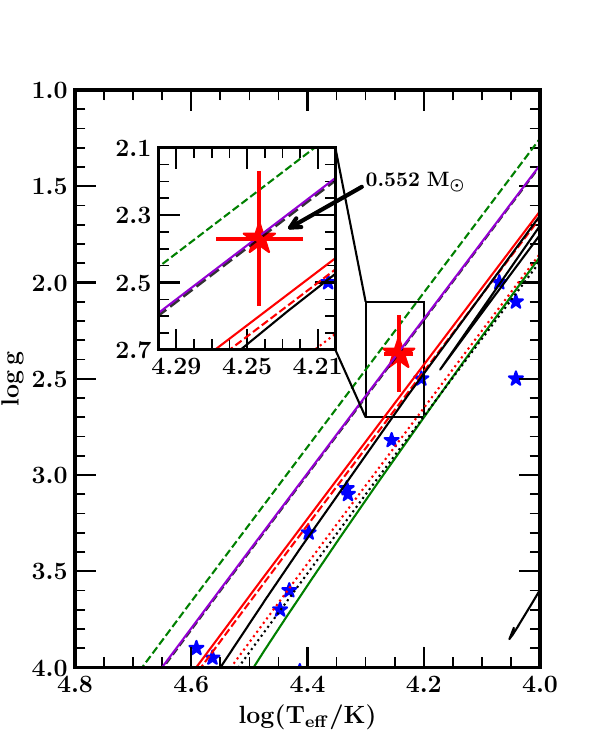}
    \caption{The log \Teff\ versus log L (left panel) and log \Teff\ versus $\log g$ (right panel) plots display PAGB stars that have been observed in GCs. The red asterisk indicates the newly discovered PAGB star in E3, and the blue asterisks indicate PAGB stars found in other GCs. In the upper left panel, the H-burning PAGB evolutionary tracks of \citet{MillerBertolami2016} and \citet{Moehler2019} with metallicity $Z = 0.001$ are shown as dashed and solid lines, respectively, and for different initial and HB masses (as shown in the legend at right). The PAGB tracks of \citet{Moehler2019} consider an initial mass of 0.85 \Msun\ for all the evolutionary tracks. The lower left panel shows the post-RGB He-burning evolutionary tracks from \citet{Bloecker1995} (green dotted line), \citet{Lei2015} (red dotted line), and \citet{Driebe1998} (black dotted line). The grey shaded region indicates a \Teff\ range of 16\,000 $-$ 19\,000 K. The current mass for each evolutionary track at $\sim$18\,000 K is shown with an arrow of the same colour and style as the evolutionary track. In the right panel, PAGB stars are shown on the log \Teff\ versus $\log g$ plane. The inset shows the zoomed-in view around the newly identified PAGB star.}
    
    \label{fig:HR-pAGB}
\end{figure*}

In Fig.~\ref{fig:HR-pAGB} we show all the previously observed PAGB stars (blue asterisks) in GCs with known \Teff, \Lbol, and $\log g$ values. The newly identified PAGB star in this paper is shown as a red asterisk. We find that the luminosity of the identified PAGB star is similar to most of the GC PAGB stars \citep[${\rm \log (L/L_{\odot})\sim3.25}$;][]{Dixon2024}. We show all the latest available PAGB evolutionary tracks with H-burning (upper left panel) and He-burning (lower left panel) prescriptions. The H-burning evolutionary tracks of initial masses 0.9, 1.0, and 1.25 \Msun\ (dashed red, black, and green lines, respectively) are from \citet{MillerBertolami2016}. The PAGB evolutionary tracks of \citet{Moehler2019} (solid lines) were generated from an initial mass of 0.85 \Msun, but they differ in their post-HB evolution based upon their location and mass on the HB phase. We find that the luminosity of evolutionary tracks varies upon their initial and/or final mass. The current mass of the identified PAGB star would be $\sim$0.51 \Msun\ based upon the H-burning evolutionary tracks (upper left panel of Fig.~\ref{fig:HR-pAGB}).

The derived $\log g$ value of the star is lower than the other PAGB stars observed in Galactic GCs (right panel of Fig.~\ref{fig:HR-pAGB}). The evolutionary tracks on the $\log g$ versus log \Teff\ plane suggest that the current mass of the identified PAGB star is $\sim$0.55 \Msun, which evolved from an initial mass of 1.0 \Msun\ \citep[][]{MillerBertolami2016}. 

Hence, various evolutionary tracks on the \Teff, \Lbol, and $\log g$ plane suggest a current mass of the PAGB star in the range 0.51 $-$ 0.55 \Msun\ which has spent $\sim$1000 yr in the PAGB phase \citep[1.0 \Msun\ track; ][]{MillerBertolami2016}. This mass range of the PAGB star indicates that the companion is also an evolved post-main sequence (of 0.5 $-$ 0.8 \Msun\ for q $\geq$ 1.0). The very close orbit of the system suggests that it would have gone into a common-envelope interaction that drastically shortened the orbital period (and ejected material into the ISM). In this case, the evolution of the star would have been interrupted compared to that of a single star. However, the star shows higher luminosity than would have resulted from interrupted evolution (black dotted line in the lower panel of Fig.~\ref{fig:HR-pAGB}, left panel), but a lower luminosity than would have resulted from a single star evolution (dashed lines in the upper panel of Fig.~\ref{fig:HR-pAGB}, left panel). 

Recently, \citet{Davis2021} showed that PAGB stars are preferentially found in metal-poor clusters with blue or very blue HB stars. On the other hand, the identified PAGB star is in E3, a relatively metal-rich cluster containing red HB stars, which is one of the few cases of metal-rich clusters that contain PAGB stars (such as the famous blue star in 47 Tucanae). However, since the PAGB in E3  has a similar luminosity to the PAGB stars in metal-poor GCs, a binary configuration better supports its presence in a metal-rich and red HB cluster. 

\section{Evidence of third dredge-up in the PAGB star} \label{sec:3DU}

PAGB stars, which undergo thermal pulses at the AGB-tip, bring up the processed materials to the surface (3DU process); as a result, the enhanced C, O, and other s-process elements could be found on the surface of PAGB stars. Recently, \citet{Dixon2024} studied the CNO abundances of 11 PAGB stars observed in the Galactic GCs and found 4 of them have enhanced CNO abundances compared to the CNO abundance of their respective host cluster. They concluded that the increased C and O abundances during their 3DU enhanced the overall CNO abundance in those PAGB stars. 

\begin{figure}
    \centering
    \includegraphics[width=0.49\columnwidth]{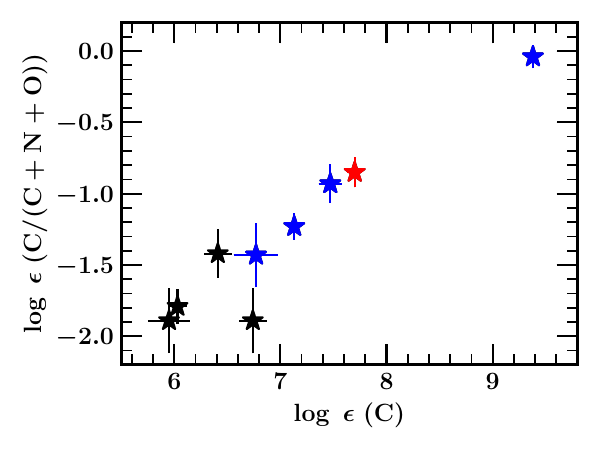}
    \includegraphics[width=0.49\columnwidth]{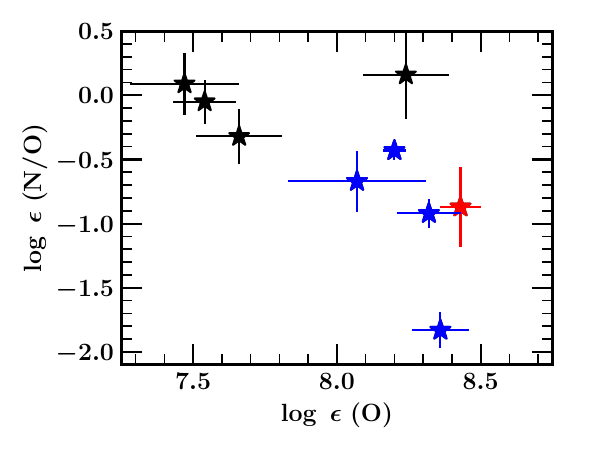}
    \caption{Comparison between abundance ratios of PAGB star of E3 and other GCs. PAGB stars with 3DU evidence are shown as blue asterisks; those with no 3DU evidence are shown as black asterisks. The newly identified PAGB star of E3 is shown as a red asterisk. The left panel shows C/CNO versus C abundance plot, whereas the right panel shows N/O vs O abundance plot.} 
    \label{fig:abundance_pAGB}
\end{figure}

In Fig.~\ref{fig:abundance_pAGB} we plot the abundance ratios of  C/CNO  versus C (left panel) and N/O versus O (right panel) abundances of the eight PAGB stars for which C abundances were reported in \citet{Dixon2024} (blue asterisks for the PAGB stars having 3DU and black asterisks for those having no sign of 3DU) along with the abundances of the identified PAGB star in this paper (red asterisk). We find that the identified PAGB star has similar C and O abundances to those PAGB stars of other GCs that have undergone the 3DU. This suggests the discovered PAGB star also went to the 3DU while on its AGB evolution. 

\section{Conclusion} \label{sec: conclusion}
We explore all the photometric and spectroscopic details of the new hot PAGB star discovered in GC E3. The proper motion and parallax of the star from Gaia DR3 support its membership, and the location of the star in the absolute CMD suggests that the star is in the PAGB phase. The radial velocities show a variation of $\sim$6 \kms\ between the two epochs, suggesting a binary nature of the star. A simulation of possible binary systems with the observed RV values of the PAGB star and the cluster mean RV suggest that the binary period of the star is around 39 days (51\% of models) or 18 days (16\% of models). The abundances were derived for the He, C, N, O, Ne, Al, Si, S, and Fe lines. The metallicity of the star ([Fe/H]=$-$0.7 dex) agrees with the cluster metallicity. We find the \Teff, $\log g$, ${\rm \log L}$, and radius of the PAGB star to be 17\,500$\pm$1\,000 K, 2.37$\pm$0.20, 3.33$\pm$0.03, and 4.61$\pm$0.14 \Rsun, respectively, using SED fitting of 30 photometric fluxes from UV to IR bands. A comparison of various PAGB evolutionary tracks with the observed star on the H-R diagram suggests that the current mass of the  PAGB star is in the range 0.51$-$0.55 \Msun. A comparison of the observed C, N, and O abundances of the star with the abundances of the PAGB stars of other GCs that show 3DU suggest that the PAGB star has undergone the 3DU and enriched its C and O abundances. Future multi-epoch spectroscopic observations are necessary to constrain the period and better understand the binary nature of the identified hot PAGB star.

\begin{acknowledgements}
    We thank the referee for his/her constructive comments. We thank Prof. M. Parthasarathy for his valuable discussion during the initial phase of the project. The CHRION spectra were obtained with the help of the National Science Center (NCN), Poland, using grant no. 2021/41/N/ST9/02746. AM is also supported by this NCN grant. ACP and SP acknowledge the support of the Indian Space Research Organisation (ISRO) under the AstroSat archival Data utilization program (No. DS\_2B-13013(2)/1/2022-Sec.2). MG was partially supported by the NCN through the grant UMO-2021/41/B/ST9/01191. DKO acknowledges the support of the Department of Atomic Energy, Government of India, under Project Identification No. RTI 4002. The research work at the Physical Research Laboratory is funded by the Department of Space, Government of India. This publication uses the data from the \mbox{{\em AstroSat}} mission of the ISRO, archived at the Indian Space Science Data Center (ISSDC).
\end{acknowledgements}

%
%
\bibliographystyle{aa}
\bibliography{E3bib}
    

\begin{appendix}

\section{Cluster membership analysis using Gaia DR3}\label{appendix: gaia}

The GC catalogue of Gaia EDR3 (GCG21; \citealt{Vasiliev2021}) provides information about the source positions (RA, DEC), proper motions (PMRA, PMDEC), parallax,\footnote{The catalogue provides zero point corrected parallax values, as suggested by \citet{Lindegren2021}.} cluster membership probability, and photometric magnitudes in G, BP, and RP filters for sources observed in the 170 Galactic GCs. We found 479  Gaia observed sources of E3 within the UVIT FoV having cluster membership probability $>$ 90\%. We cross-matched the eight UVIT observed sources with the Gaia catalogue and found only one source to be a cluster member. The UVIT observed cluster member source has a membership probability of 99\% and lies well within the half-light radius of the cluster. For parallax information, we selected only those cluster member sources with good parallax (S/N > 5) and found only 8 out of 479 cluster member sources. The mean proper motions and parallax of the cluster estimated by \citet{Vasiliev2021} are PMRA (${\rm \mu_\alpha cos \delta}$) $= -2.727 \pm 0.027$ (mas/yr), PMDEC (${\rm \mu_\delta}$) = 7.083 $\pm$ 0.027 (mas/yr), and parallax ($\Bar{\omega}$) =  0.146 $\pm$ 0.013, respectively. We find that the proper motion and the parallax values of the UVIT observed cluster member source are in good agreement with the Gaia observed cluster member sources (Fig.~\ref{fig:gaia_pm}). The details of the UVIT and the Gaia observations of the source are given in Table \ref{tab:pAGB_details}.

In the right panel of Fig.~\ref{fig:gaia_pm}, we show the optical CMD of the cluster member sources (grey dots) in the absolute magnitude plane. The colours and magnitudes of the  Gaia filters are dereddened using E(B$-$V) = 0.29 \citep{Schlafly2011} and the \citet{cardeli1989} extinction law. The Gaia G magnitudes are scaled to the absolute magnitudes using the distance modulus, m$-$M = 14.50 \citep{Baumgardt2021}. The UVIT observed source (blue asterisk) is $\sim$2.0 mag brighter than the other cluster member sources and appears bluer in colour (BP$-$RP $= -$0.2 mag), which is the location of the blue evolved stars of GCs \citep{Zinn1972, Moehler2019, Bond2021}. We marked various regions of PAGB stars, defined by \citet{Bond2021} as red-PAGB (rPAGB), yellow-PAGB (yPAGB), blue-PAGB (bPAGB), and AGB-manqu\'e stars, on the Gaia BP$-$RP versus G CMD (right panel of Fig.~\ref{fig:gaia_pm}). The UVIT observed cluster member source can be found in the bPAGB region. Hence, we confirm that the source is an evolved PAGB star and a bona fide member of the GC E3.

\section{Reduction of CHIRON spectra of the PAGB star}\label{appendix: A}

We reduced the spectra with the pipeline developed at Yale University \citep{Tokovinin2013}. Wavelength calibrations were performed using the ThAr lamp exposures taken during the scheduled calibration runs. We applied barycentric corrections separately using calculations from IRAF with the \textit{bcvcor} task. We calculated the signal-to-noise ratio (S/N) of the two spectra using the  \texttt{snr}\footnote{\href{specutils.readthedocs.io/en/stable/api/specutils.analysis.snr.html}{https://specutils.readthedocs.io/en/stable/analysis.html}} module in \textsc{specutils}. We found a S/N of 40.6 and 25.6 for the spectra at the first and second epochs, respectively.

 \begin{table}
    \caption{Details of the spectroscopic observations of the PAGB star.} 
    \label{tab:pagb_spec}
    \centering
    \begin{tabular}{cccccc}
    \hline
    Epoch   & Date of Observation & Exposure Time & S/N \\ \hline
       1  & 17 February 2023 & 4 $\times$ 1600 sec. & 40.6 \\
       2 & 26 March 2023 &  4 $\times$ 1600 sec. & 25.6 \\
       \hline
    \end{tabular}
\end{table}

The optical spectrum of the star in the spectral ranges 4505$-$6602 \AA\ shows about 50 stellar absorption lines. The identification of the lines is based on the National Institute of Standards and Technology (NIST) Atomic Spectra Database.\footnote{\href{https://www.nist.gov/pml/atomic-spectra-database}{https://www.nist.gov/pml/atomic-spectra-database}}

Absorption lines of neutral species including \ion{H}{i} (H$\alpha$ and H$\beta$),
\ion{He}{i}, \ion{C}{I}, \ion{N}{I}, \ion{O}{I},  and \ion{Ne}{I}
were identified. Singly ionised species including \ion{C}{II},
\ion{N}{II}, \ion{O}{II}, \ion{Si}{II}, \ion{S}{II},
\ion{Mg}{II}, and \ion{Fe}{II} were detected. Higher ionisation is seen in
\ion{Al}{III}, \ion{Fe}{III}, \ion{Si}{III}, and \ion{S}{III}.

The spectrum of the star also contains absorption features that have interstellar origin. There is a \ion{Na}{I} doublet ($\lambda$5889.951, 5895.924) and several very weak diffuse interstellar bands (DIBs). Emission lines are not detected in the spectrum of the star, in contrast to many hot post-AGB stars, whose spectra consist of two components: the star's absorption spectrum and an overlain emission spectrum from a low-excitation gas envelope \citep{Mello2012}. Comparing the spectra at the two epochs, we find static \ion{O}{I} emission lines at 5577 \AA, 6300 \AA, and 6364 \AA, which are most probably due to air-glow \citep{OIatm1998}, and therefore we did not consider them for our analysis. The complete continuum-normalised spectrum of the star in the spectral ranges 4\,505$-$6\,602 \AA\ is presented at \href{http://lnfm1.sai.msu.ru/~davnv/E3/atlas_E3.pdf}{http://lnfm1.sai.msu.ru/$\sim$davnv/E3/atlas\_E3.pdf}.

\section{RV estimation from the spectra of the PAGB star}\label{appendix: B}

\begin{figure}
    \centering
    \includegraphics[width=\columnwidth]{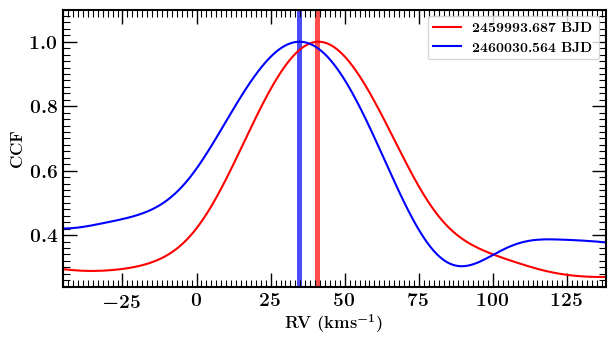}
    \caption{1D projections of the \textsc{todcor} cross-correlation functions (CCFs) for the spectra observed at two epochs. The vertical  shaded regions represent the errors in the observation of the RVs. The first epoch (red) has a RV of 40.20 $\pm$ 0.57 $\mathrm{km}\mathrm{s}^{-1}$ and the second epoch (blue) has a RV of 34.66 $\pm$ 0.85 $\mathrm{km}\mathrm{s}^{-1}$.}
    \label{fig:ccftodcor}
\end{figure}

The RVs were measured on prominent absorption lines, including He lines, but avoiding H$\alpha$ and H$\beta$ as they had broad absorption lines (FWHM $>100$~\kms) affecting a precise RV estimate. We used a template with \Teff\ = 20\,000~K, $\log g$ = 4.5 dex, and solar metallicity to create the cross-correlation functions (CCF; Fig.~\ref{fig:ccftodcor}) using the {\sc todcor} technique \citep{Zuckermazeh1994}. The uncertainties were calculated using a bootstrap procedure \citep{Helminiak2012}, which is sensitive to the S/N of a component and velocity of rotation.  The RVs were estimated as 40.20$\pm$0.57 \kms\ and 34.66$\pm$0.85 \kms\ for the spectra observed at the first and second epochs, respectively. We also checked for the consistency of RV measurements using broadening functions (BFs; \citealt{Rucinski1999}). The RVs were calculated using a rotational broadening model described in \citet{Moharana2023}. The BF-RVs were consistent with the {\sc todcor}-RVs within the errors. The final RVs were taken from {\sc todcor} due to their high precision. 

\FloatBarrier

\section{Simulation of binary configurations} \label{appendix: C}

\noindent To check the possible binary configurations of the identified PAGB star, we made a grid of models constrained by the observed radial velocities. We used the binary modelling code \textsc{phoebe2} (\citealt{Conroy2020} and references therein). Since we had two epochs of RV observations only, we considered the following assumptions for a simplified modelling: 

\begin{enumerate}
     \item The binary is edge-on. The inclination of the orbital plane is fixed at 87 degrees, and the orbital period is constrained to be less than 1000 days. For a wider orbit, we expect a high renormalised unit weight error (RUWE) parameter in Gaia DR3; however, the RUWE value of the PAGB star is 1.022 (< 1.4). 

    \item Models with orbits smaller than the radius of the PAGB star (4.61 \Rsun) or reaching the Roche limit will be rejected. We assume the secondary radius to be 1.00 \Rsun.  

    \item The centre of mass (COM) or the gamma velocity of the binary is fixed at 12.6 km/s (the cluster mean). 
\end{enumerate} 

\noindent The model generation followed these steps:
\begin{enumerate}
    \item Setting up an initial model with random draws from a set of masses (M$_{\rm pri}$ = 0.2, 0.5, 0.6, and 0.8 \Msun), eccentricities (e = 0.0, 0.2, 0.5, and 0.8), mass ratios (q = 0.5, 1.0, and 1.5), and binary periods (P; randomly between 0$-$90 days). The mass ratios and binary periods are guess values that are needed in order to initialise the \textsc{phoebe2} set-up. We generated an initial set-up of 480 models. 
    
    \item We apply the constraints and then q, P, and time of periastron passage ($T_p$) to best fit the observed RV values. The optimisation was done using the Nelder-Mead method \citep{neldermead} for 500 iterations.  
    
    \item The optimisation discarded a few models if they failed to adhere to any pre-defined constraints. This gave us the final 396 models with possible q and P values. 
\end{enumerate}

\begin{figure*}
    \centering
    \includegraphics[width=0.68\textwidth]{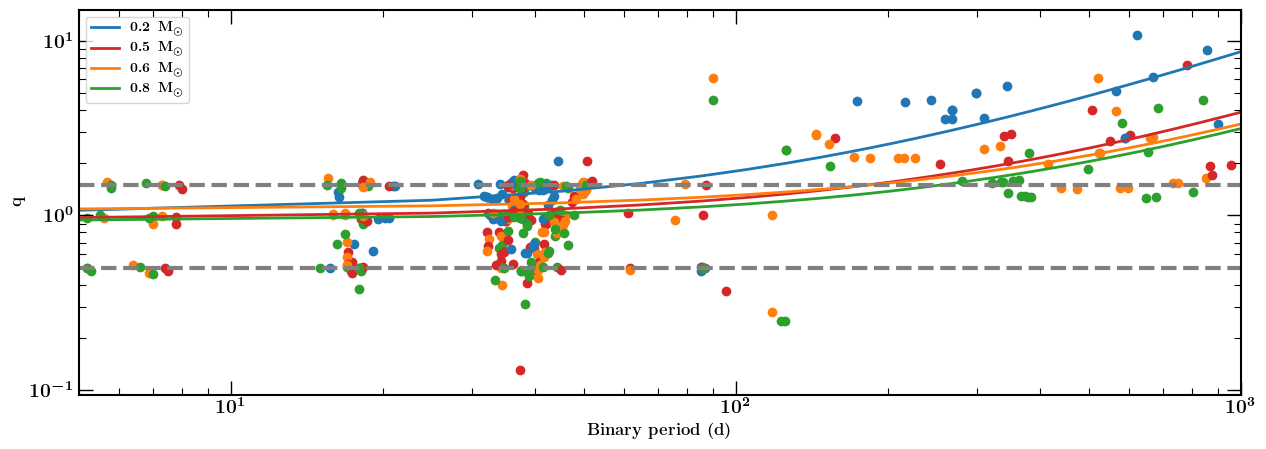}
    \includegraphics[width=0.2925\textwidth]{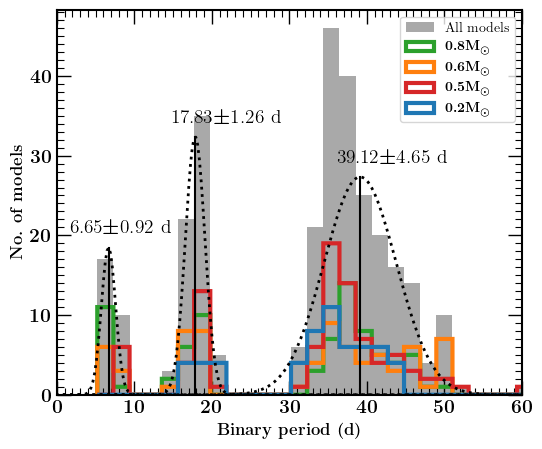}
    \caption{ Orbital period and mass ratio values obtained from the best-fitting binary models. Left panel: Distribution of mass ratio (q) and orbital periods (P) with primary masses 0.2, 0.5, 0.6, and 0.8 \Msun\ (blue, red, orange, and green dots, respectively). The solid lines are linear fits of the q and P values for the separate primary masses considered. The grey dashed lines denote the range of favourable q values for the system. 56.6 \% of the models have mass ratios between 0.5 and 1.5. Right panel: Distribution of the period of the best-fit models. We see three significant period distributions at $\sim$6 d, $\sim$18 d, and $\sim$39 d, with the 39 d distribution being more probable (51\%).}
    \label{fig:period_q}
\end{figure*}

The possible q and P values for our final binary models are shown in the left panel of Fig.~\ref{fig:period_q}. The best-fit binary periods are clustered around three values: 39.12$\pm$4.65 days (51.26 \% of the models), 17.83$\pm$1.26 days (16.4 \%), and 6.65$\pm$0.92 days (6.8\%) (right panel of Fig.~\ref{fig:period_q}). The mass ratio, q, varies between 0.2 and 1.5 (56 \%; grey dashed lines); however, the average values are around 1.0 for periods < 100 days (blue, red, orange, and green solid lines for masses 0.2, 0.5, 0.6, and 0.8 \Msun, respectively). We do not see any favourable eccentricity, but it is hard to constrain it with two epochs of observations. 

\begin{figure*}
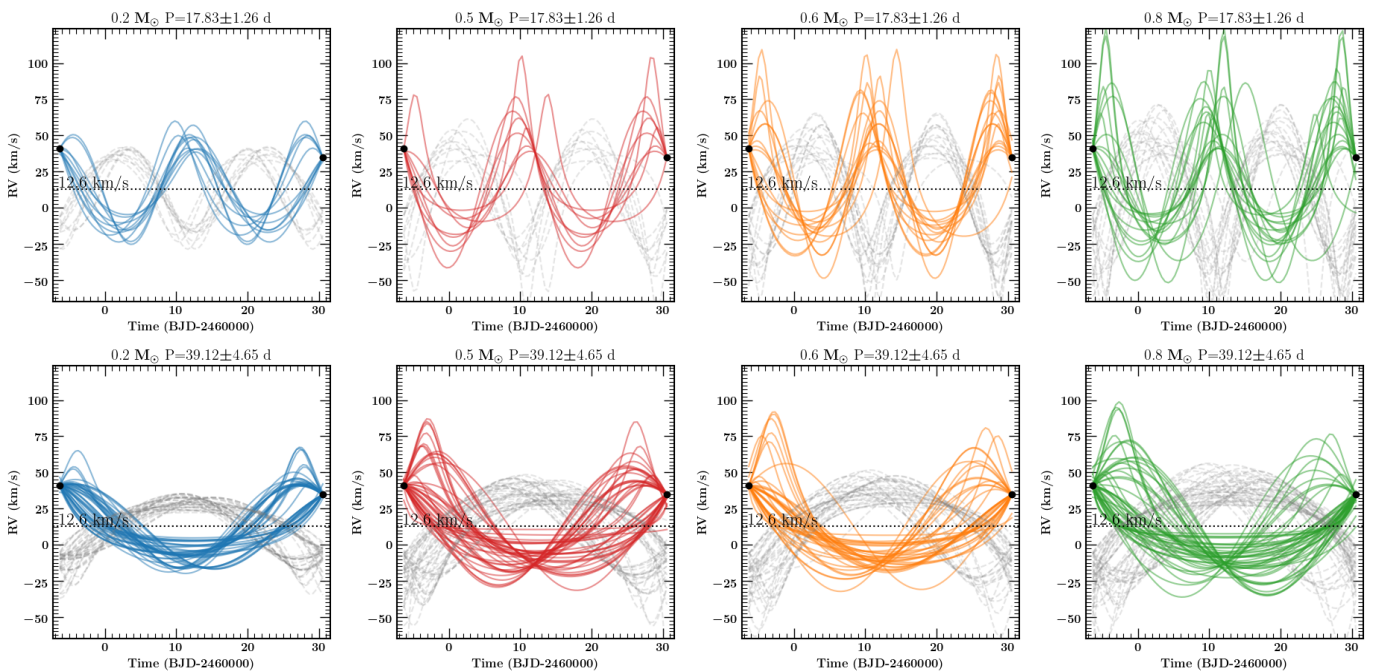

    \centering
    \includegraphics[width=0.24\textwidth]{0.2 M_17.83d_RV.png}
    \includegraphics[width=0.24\textwidth]{0.5 M_17.83d_RV.png}
    \includegraphics[width=0.24\textwidth]{0.6 M_17.83d_RV.png}
    \includegraphics[width=0.24\textwidth]{0.8 M_17.83d_RV.png}
    
    \includegraphics[width=0.24\textwidth]{0.2 M_39.12d_RV.png} 
    \includegraphics[width=0.24\textwidth]{0.5 M_39.12d_RV.png} 
    \includegraphics[width=0.24\textwidth]{0.6 M_39.12d_RV.png} 
    \includegraphics[width=0.24\textwidth]{0.8 M_39.12d_RV.png}
 
    \caption{Possible Keplerian models for radial velocity variations constrained to the observations (black dots).  The models are for PAGB mass 0.2 \Msun\ (blue), 0.5 \Msun\ (red), 0.6 \Msun\ (orange), and 0.8 \Msun\ (green) around the orbital periods 17.83$\pm$1.26 days (upper panels) and 39.12$\pm$4.65 days (lower panels). The dashed grey curves represent the possible RV of the secondary to maintain the COM velocity as 12.6 km/s (black dotted line). }
    \label{fig:RV_variation}
\end{figure*}

In Fig.~\ref{fig:RV_variation} we provide the possible RV values for the primary (coloured solid lines) and companion stars (grey solid lines) for a period of 17.83 days (upper panels) and 39.12 days (lower panels). The maximum RV can reach $\sim$100 km/s for the PAGB star with masses 0.6 \Msun\ (period = 17.83 days) and 0.8 \Msun\ (period = 39.12 days) (Fig.~\ref{fig:RV_variation}).

\FloatBarrier

\section{Atmospheric parameter estimation from the high-resolution spectra}\label{appendix: D}

To derive the parameters of the stellar atmosphere from the  CHIRON spectra, we used the {\sc tlusty} model grid BSTAR2006 \citep{Lanz2007} and accompanied programs for calculating synthetic spectra ({\sc synspec}, {\sc rotin}; see \citealt{Hubeny2021}).

Synthetic profiles of the Balmer lines fit the observations only at a certain relation between \Teff\ and $\log g.$ In the case of H$\alpha$ and H$\beta$ lines, these relations are similar in shape, but shifted relative to each other: $\log g_{\beta}(\rm T_{ eff})=\log g_{\alpha}(\rm T_{ eff})-0.4.$ The higher members of the Balmer series do not fall in the observed range (4505$-$6602 \AA). Therefore, as the final relation for $\log g$--\Teff, we took the average between H$\alpha$ and H$\beta$ with an uncertainty of 0.2 dex.

\begin{figure}
\begin{center}
\includegraphics[scale=0.65]{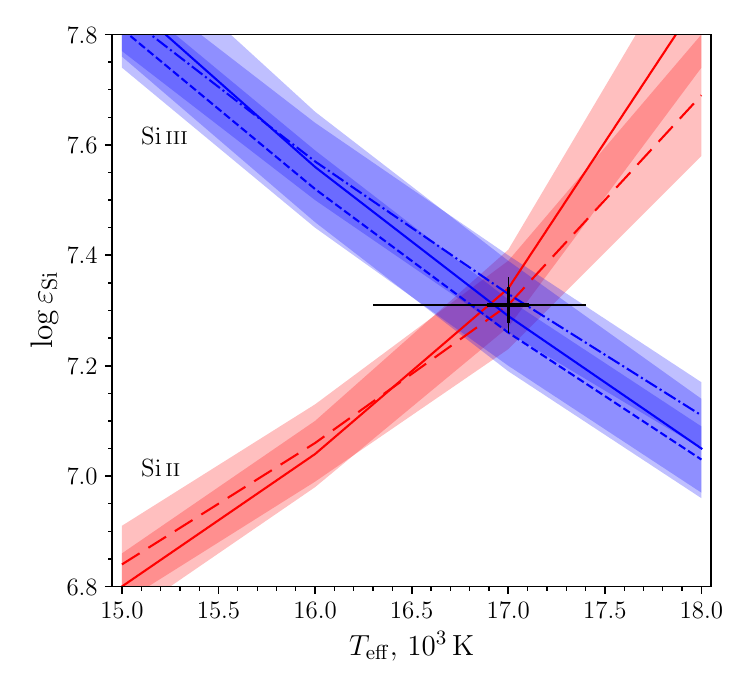}
\end{center}
\caption{Silicon ionisation balance. Red   is for \ion{Si}{ii};  the solid line is for \ion{Si}{ii}\,$\lambda$6347, the dashed line is for \ion{Si}{ii}\,$\lambda$6371. Blue   is for \ion{Si}{iii};   the solid, dashed, dash-dotted lines are for \ion{Si}{iii}\,$\lambda\lambda$4552, 4567, 4572, respectively. The shaded regions show the estimated uncertainties for each line. }\label{tlgg}
\end{figure}

The observations allowed us to construct a second relation involving the ionisation balance between \ion{Si}{ii} and \ion{Si}{iii}, whose lines are quite strong in the spectrum. We adjusted the silicon abundance ($\varepsilon_{\rm Si}$) for each line of \ion{Si}{ii}\,$\lambda$6347, 6371 and \ion{Si}{iii}\,$\lambda$4552, 4567, 4572 for a set of pairs (\Teff, $\log g$) along the found relation $\log g$(\Teff). For the true values of \Teff\ and $\log g$, the abundances $\varepsilon_{\rm Si}$ derived from \ion{Si}{ii} and \ion{Si}{iii} must coincide. We find that the coincidence of $\varepsilon_{\rm Si}$ over all lines \ion{Si}{iii}/\ion{Si}{ii} occurs at \Teff = $16950\pm100$ K, $\log \varepsilon_{\rm Si}=7.31\pm0.03$, and $\xi=1.5\pm0.5$\kms\ (see Fig.~\ref{tlgg}). A big uncertainty of the corresponding value of $\log g=2.37\pm0.20$ leads to an increase in the temperature uncertainty, making it asymmetric $\rm T_{eff}=17\,000_{-600}^{+300}$ K. All calculations were made with the BSTAR2006 models with scaled solar abundances $Z/Z_\odot=1/2$ (BL models) and $\xi=2$\kms. We also redid all the calculations with the $Z/Z_\odot=1/5$ (BS models) and $\xi=2$\kms\ to know the effect of variations in the chemical composition in the models. We find that it does not affect the relation $\log g - \rm T_{eff},$ but it reduces \Teff\ by 300 K in the diagram $\rm T_{eff} - \log \varepsilon_{\rm Si}$ and increases $\log \varepsilon_{\rm Si}$ by 0.04 dex. Thus, the final results are $\rm T_{eff}=17\,000_{-700}^{+400}$~K, $\log g = 2.37\pm0.20$, and  $\log \varepsilon_{\rm Si}=7.31\pm0.05.$

\FloatBarrier

\section{Chemical composition of the individual elements}\label{appendix: E}

\begin{table}[h]
\caption{Element abundances determined for the identified PAGB star. The abundances are estimated as $\log \varepsilon=12+\log(n_\text{X}/n_\text{H})$ and [X/H]=$\log(n_\text{X}/n_\text{H})-log(n_{\text{X}\odot}/n_{\text{H}\odot})$. The solar values $\log \varepsilon_{\odot}$ are from \citet{Asplund2009}; $\sigma_L$ is the standard deviation; $\overline{\sigma}$ is the final uncertainty, accounting for inaccuracies in the model parameters; N is the number of averaged lines.}
\label{tab:abund}
\begin{center}
\begin{tabular}{lcccccc}
\hline \hline
        & $\log \varepsilon_{\odot}$   & $\log \varepsilon$ & [X/H]& $\sigma_L$& $\overline{\sigma}$ &N  \\
\hline
He      &  10.93  &  11.00    &$+$0.07    &   0.03    &  0.08     & 4 \\
C       &   8.43  &   7.70    &$-$0.73   &   0.03    &  0.07     & 2 \\
N       &   7.83  &   7.56    &$-$0.27   &   0.29    &  0.3      & 9 \\
O       &   8.69  &   8.43    &$-$0.26   &   0.04    &  0.07     & 4 \\
Ne      &   7.84  &   7.57    &$-$0.27   &   0.02    &  0.03     & 3 \\
Al      &   6.45  &   5.61    &$-$0.84   &   0.18    &  0.19     & 2 \\
Si      &   7.51  &   7.31    &$-$0.20   &   0.03    &  0.05     & 5 \\
S       &   7.12  &  $<$ 6.59   & $< -$0.53  &   $-$      &   $-$      & 6 \\
Fe      &   7.50  &   6.80    &$-$0.70   &   $-$      &  0.2      & 1 \\
\hline
\end{tabular}
\end{center}
\end{table}

\begin{figure}
    \begin{center}
    \includegraphics[scale=0.50]{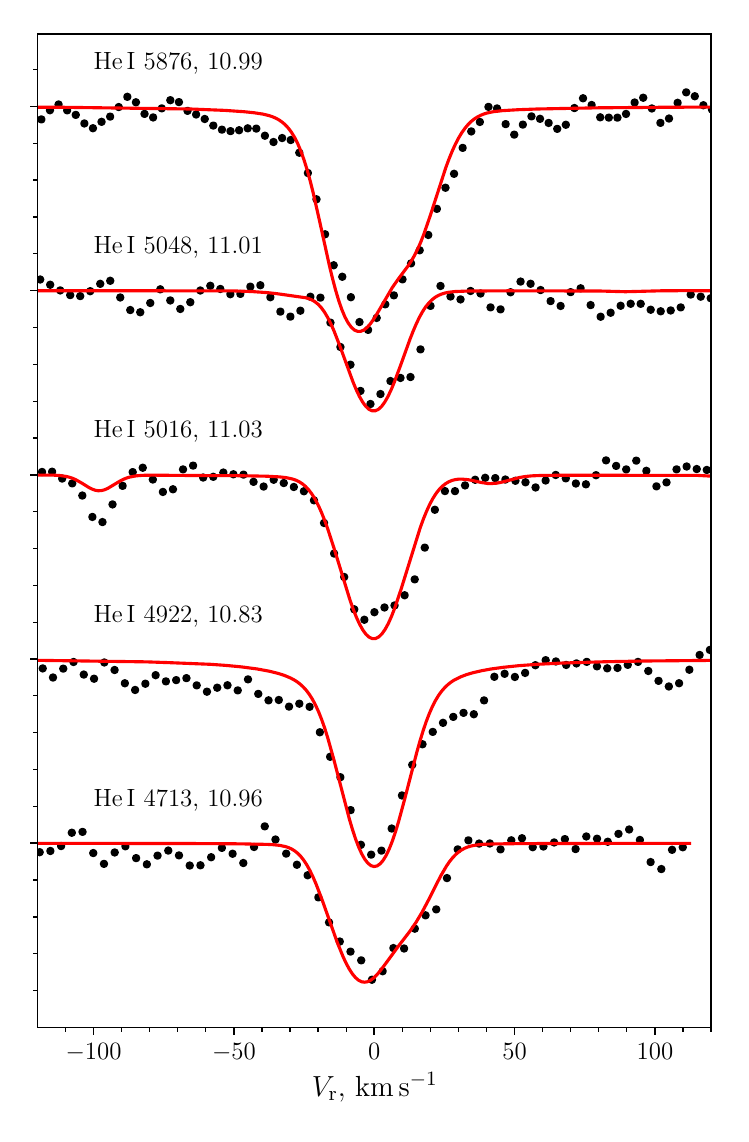}
    \end{center}
    \caption{Observed profiles of \ion{He}{i}\ lines (dots) are superposed on the model spectra (red lines). The numbers represent the values of $\log \varepsilon_{\rm He}$.}
    \label{HeI}
\end{figure}

{\it Helium.} \ion{He}{i}\,$\lambda\lambda$4713, 4922, 5016, 5048, 5876 lines are considered. We removed the  \ion{He}{i}\,$\lambda$4922 line from the averaging, because it is blended with some broad absorption feature and shows $\log \varepsilon_{\rm He}$ lower by 0.2 dex in comparison with the other lines. The rest of the lines agree well with each other. Figure~\ref{HeI} shows the \ion{He}{i}\ line profiles and fragments of the synthetic spectrum.

\begin{figure}
\begin{center}
\includegraphics[scale=0.50]{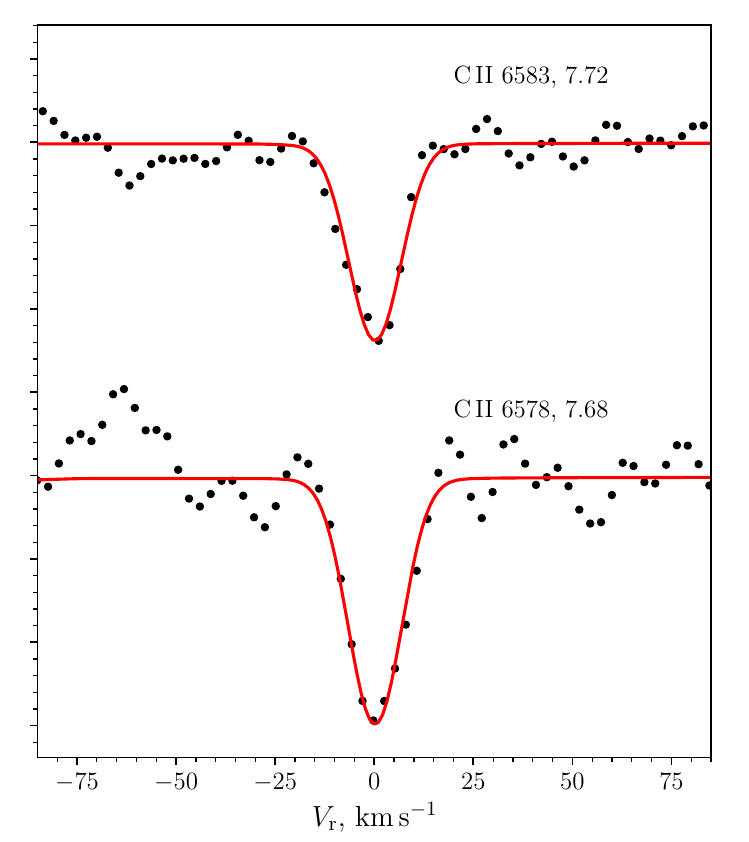}
\end{center}
\caption{\ion{C}{ii}\ line profiles. The
plot symbols are the same as in Fig.~\ref{HeI}. The numbers represent the values of $\log \varepsilon_{\rm C}$.}
\label{CII}
\end{figure}

{\it Carbon.} The abundance $\varepsilon_{\rm C}$ is derived from two strong lines \ion{C}{ii}\,$\lambda \lambda$ 6578, 6583. Figure~\ref{CII} shows the \ion{C}{ii}\ lines profiles superposed on model spectra.

\begin{figure}
    \begin{center}
    \includegraphics[scale=0.50]{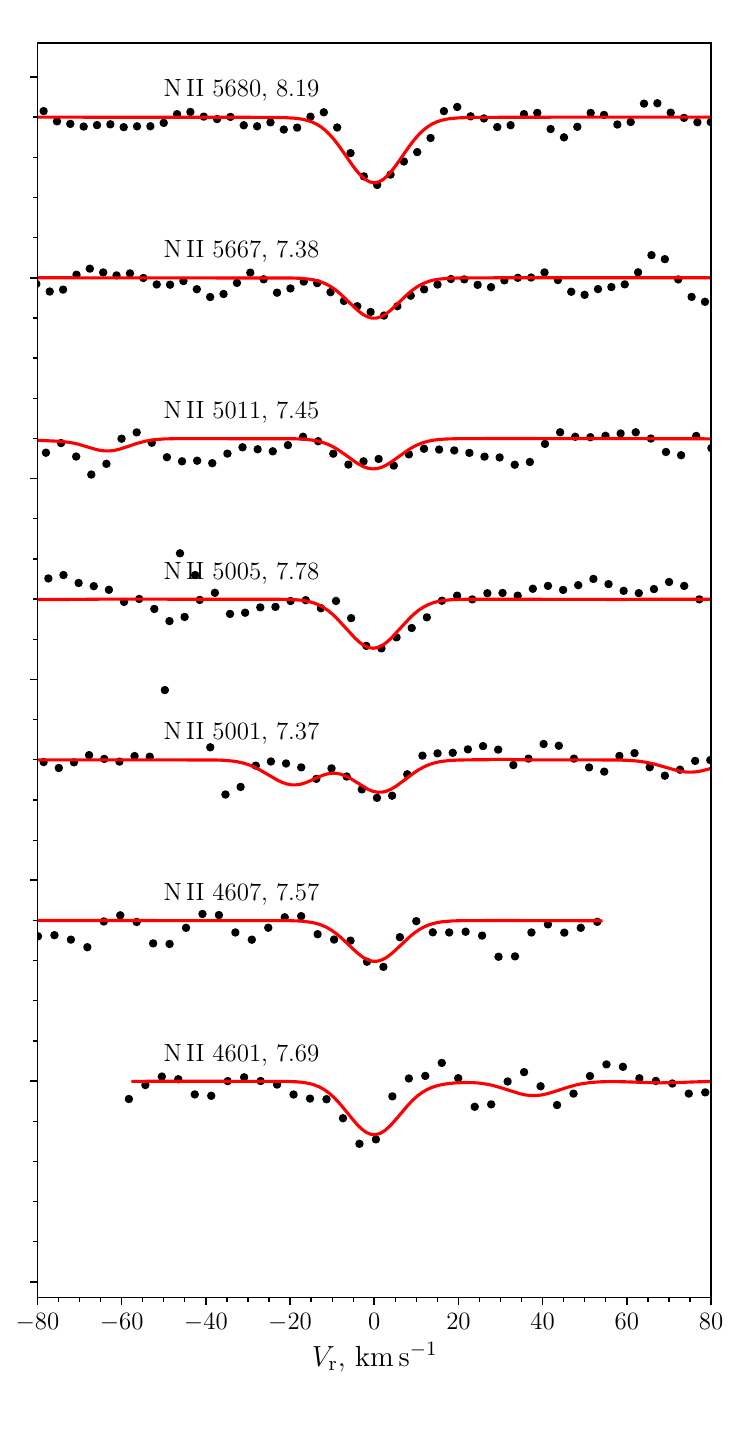}
    \end{center}
    \caption{\ion{N}{ii}\ line profiles. The
    plot symbols are the same as in Fig.~\ref{HeI}. The numbers represent the values of $\log \varepsilon_{\rm N}$.}
    \label{NII}
\end{figure}

{\it Nitrogen.} The observed spectrum contains several weak lines: \ion{N}{ii}\,$\lambda\lambda$4601, 4607, 4630, 5000, 5001, 5005, 5011, 5667, 5680. The abundances derived from the various lines do not agree well with each other; stronger lines require a higher abundance than weaker ones. It indicates a high microturbulence velocity $\xi>10$\kms, but the observed width of the lines is small and almost completely determined by the instrumental profile that puts a limit on $\xi<5$\kms.  Figure~\ref{NII} shows the \ion{N}{ii}\ line profiles superposed on model spectra.

\begin{figure}
    \begin{center}
    \includegraphics[scale=0.50]{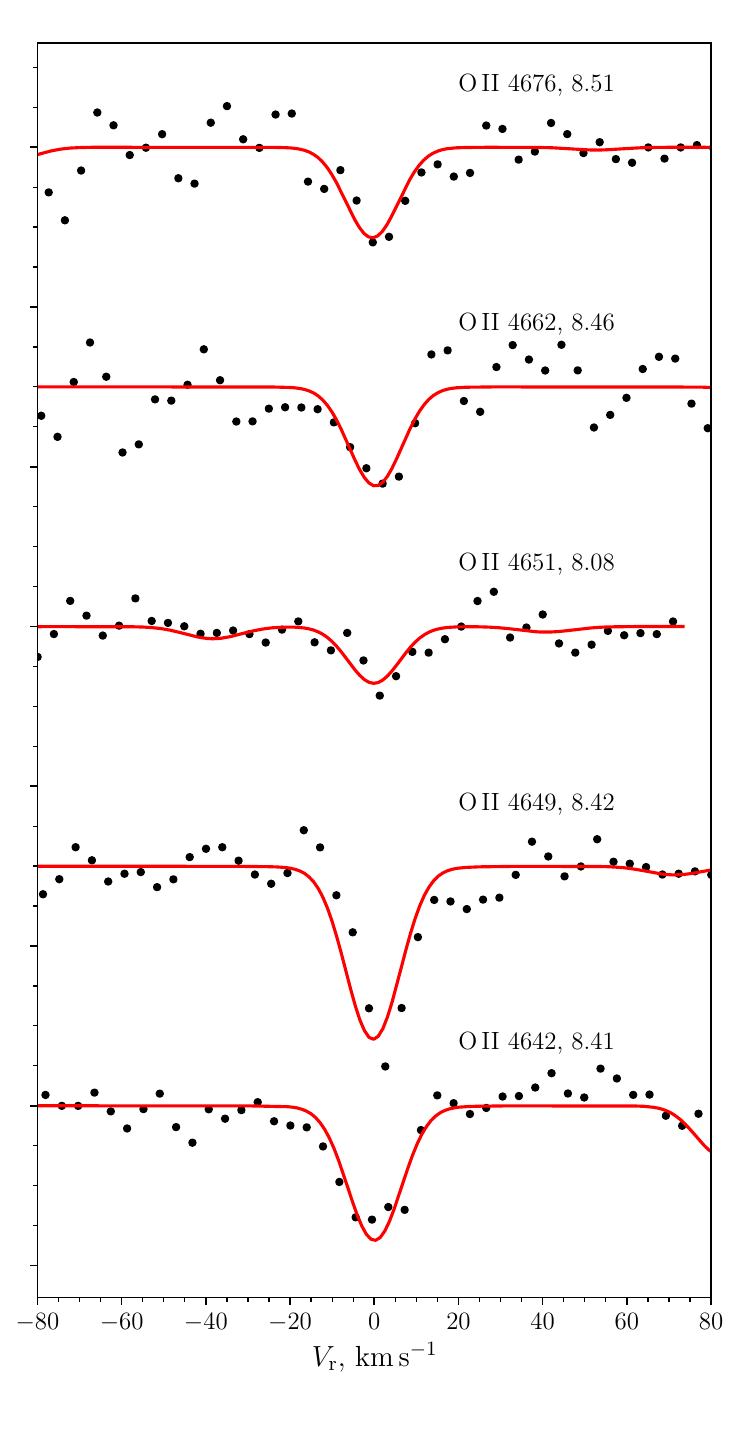}
    \end{center}
    \caption{\ion{O}{ii}\ line profiles. The
    plot symbols are the same as in Fig.~\ref{HeI}. The numbers represent the values of $\log \varepsilon_{\rm O}$.}
    \label{OII}
\end{figure}

{\it Oxygen.} A few weak lines of \ion{O}{ii} are seen in the spectrum. We used \ion{O}{ii}\,$\lambda\lambda$4642, 4649, 4662, and 4676. Figure~\ref{OII} shows the \ion{O}{ii}\ lines profiles superposed on model spectra.

{\it Neon.} The abundance $\varepsilon_{\rm Ne}$ is derived from \ion{Ne}{i}\,$\lambda\lambda$6143, 6402, and  6507, which agree well with each other. There are two weaker lines \ion{Ne}{i}\,$\lambda\lambda$6266, 6599, but they show a $\log \varepsilon_{\rm Ne}$ value that is  0.3 dex lower and  0.3 dex higher than the rest of the lines.

{\it Aluminium.} Two lines are used: \ion{Al}{iii}\,$\lambda\lambda$5697, 5723.

{\it Silicon.} The silicon abundance is determined simultaneously with the atmospheric parameters (see above).

{\it Sulphur.} We considered lines \ion{S}{ii}\,$\lambda\lambda$5014, 5032, 5432, 5454, 5640, and  5647, although only the lines $\lambda\lambda$5014, 5640, and  5647 are reliably present in the observed spectrum. These lines give $\log \varepsilon_{\rm S}=$ 6.51, 6.26, and 6.59, respectively. The  \ion{S}{ii}\,$\lambda$5032 line is relatively weak, but it is in agreement with the above-mentioned $\varepsilon_{\rm S}.$ However, at this abundance the lines \ion{S}{ii}\,$\lambda\lambda$5432, 5454 must be present in the observations. The absence of these lines puts a limit $\log \varepsilon_{\rm S}\lesssim5.5.$ In any case, there is a sulphur deficiency $\log \varepsilon_S<6.59$ ([S/H]$<-0.53$).

{\it Iron.} Only one Fe line \ion{Fe}{iii}\,$\lambda$5156 is visible in the spectrum and gives $\log \varepsilon_{\rm Fe}=6.8.$ More weaker lines \ion{Fe}{ii}\,$\lambda\lambda$5018, 5169 and \ion{Fe}{iii}\,$\lambda$5127 are predicted by the model and agree with the observations within the uncertainties at the found abundance. The Fe lines are very weak in the spectra of early-type stars, making the estimate of metallicity based on iron a very difficult task. Therefore, we must rely on the CNO abundances as metallicity proxies \citep{Mello2012}. For the star, we found that Z(CNO) is equal to 0.005, compared to the solar value of 0.012.

The final uncertainty $\overline{\sigma}$ we estimated as $\overline{\sigma}^2 = \sigma_L^2+\sigma_T^2+\sigma_g^2+\sigma_\xi^2+\sigma_M^2,$ where $\sigma_T$, $\sigma_g$, and  $\sigma_\xi$ are uncertainties in $\log \varepsilon,$ related to the uncertainties of $T_{\rm eff}$, $\log g$, and $\xi$, respectively. The quantity $\sigma_M$ is an estimate of the uncertainty associated with the difference between the real metallicity and the given metallicity of the BL models.

\FloatBarrier

\section{SED fitting of the PAGB star} \label{appendix: F}

\begin{table*}
    \caption{Telescopes and their filters used in the SED fittings of the UVIT observed PAGB star.}
    \label{tab:telescope}
    \centering
    
    \adjustbox{max width=\textwidth} {
    \begin{tabular}{ c c c  c  }
    \hline
    Telescope/Survey & Filters & Wavelength range &  Reference\\
    \hline
    UVIT & BaF2, Sapphire & 1350$-$1800 \AA &  This paper \\ 
    {\em GALEX} & FUV, NUV & 1350$-$3000 \AA & \citet{Schiavon2012} \\
    Swift/UVOT & UVW2, UVM2, UVW1 & 1600$-$4700 \AA & HEASARC Archive \\
    Gaia  & G, BP, RP & 3300$-$10600 \AA &  \citet{Vasiliev2021} \\
    CTIO-4m  & U, B, V, R, I & 3000$-$11800 \AA &  \citet{Stetson2019} \\
    APASS & B, V, g, r, i & 4300$-$7457 \AA & \citet{Henden2015} \\
    DENIS & I, J, Ks & 7861$-$21453 \AA & \citet{Denis2005} \\
    2MASS  & J, H, Ks &  12000$-$21500 \AA  &  \citet{Skrutskie2006} \\
    WISE  & W1, W2, W3, W4 &  34000$-$220000 \AA  &  \citet{Cutri2014} \\
    \hline
    \end{tabular} }
\end{table*}

\begin{figure}
    \centering
    \includegraphics[width=0.495\textwidth]{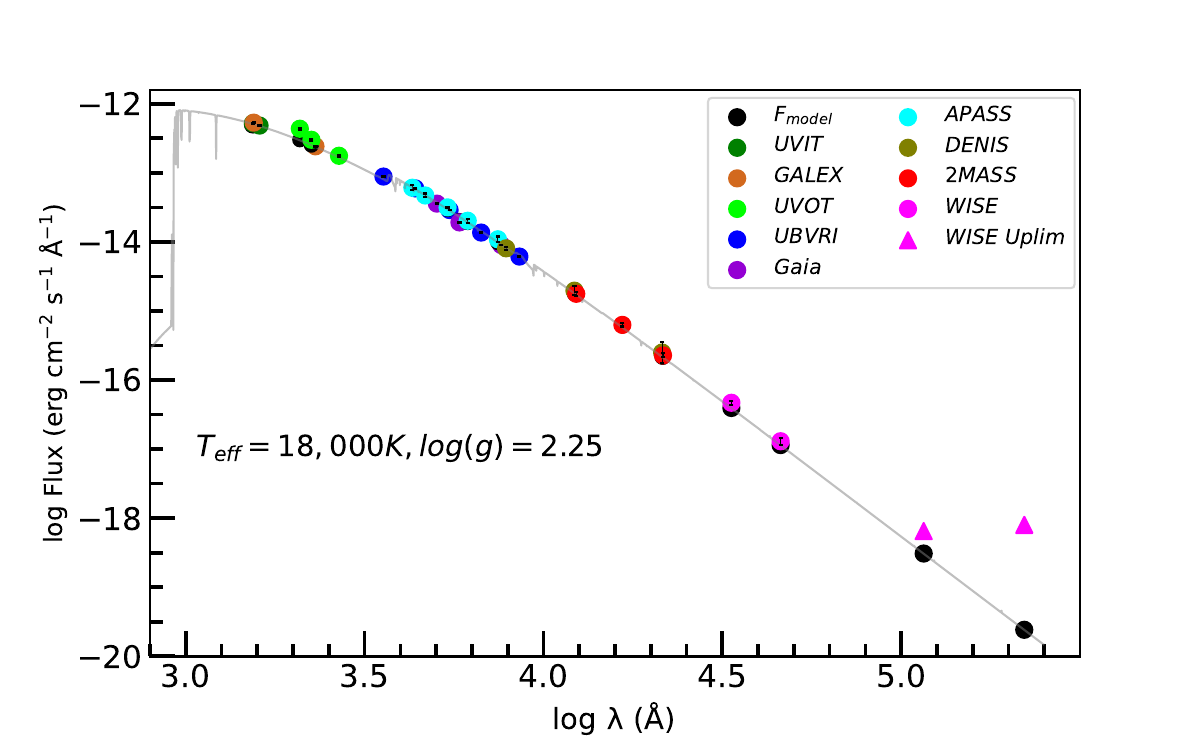}
    \caption{SED fit of observed fluxes in UVIT, GALEX, UVOT, UVBRI, Gaia DR3, APASS, DENIS, 2MASS, and WISE filters (30 photometric fluxes in total from UV to IR) for the PAGB star. The coloured solids are the observed fluxes, the black solids are the best-fitting model fluxes, and the grey solid is the best-fitting TLUSTY model grid spectra.}
    \label{fig:sed_fit}
\end{figure}

The SED fitting of the PAGB star was performed using 30 photometric fluxes observed in different filters ranging from the  UV to IR bands (Table \ref{tab:telescope}) and the theoretical fluxes from the TLUSTY non-LTE O/B stellar atmosphere model\footnote{\href{http://tlusty.oca.eu/Tlusty2002/tlusty-frames-cloudy.html}{http://tlusty.oca.eu/Tlusty2002/tlusty-frames-cloudy.html}} \citep[TLUSTY model; ][]{Hubeny1995, Lanz2007, Hubeny2021}. The SED fitting procedure was performed in VO SED Analyzer\footnote{\href{http://svo2.cab.inta-csic.es/theory/main/}{http://svo2.cab.inta-csic.es/theory/main/}} \citep[VOSA; ][]{Bayo2008}. We fitted the observed photometric fluxes with the  TLUSTY model grids and found that the best-fitting model grid has \Teff=18\,000~K and $\log g$ = 2.25. The best-fitting model and observed fluxes are shown in Fig.~\ref{fig:sed_fit}.

\end{appendix}

\end{document}